\documentclass[published]{JHEP3}
\usepackage{epsfig,multicol,bbm}
\usepackage{amsmath}
\JHEP{00(2007)000}

\title{Clausius relation and Friedmann equation in FRW universe model}

\author{Qiao-Jun Cao, Yi-Xin Chen\thanks{Corresponding},  Kai-Nan Shao\\
Zhejiang Institute of Modern Physics, Zhejiang University\\
Hangzhou 310027, P. R. China\\
E-mail: \email{caoqiaojun@gmail.com}, \email{yxchen@zimp.zju.edu.cn}, \email{shaokn@gmail.com}}

\abstract{
It has been shown that Friedmann equation of FRW universe can
be derived from the first law of thermodynamics in Einstein gravity,
Gauss-Bonnet gravity, Lovelock gravity, scalar-tensor gravity and
$f(R)$ gravity. Moreover, it was pointed out that the temperature of the
apparent horizon can be obtained using the tunneling formalism for
the corresponding observers defined by Kodama vector. In this article,
we find that the energy flux through the apparent horizon can be determined
by using the Kodama vector. 
This implies the fact that the Clausius relation and the first law of thermodynamics  associated
with the apparent horizon in FRW universe
is relative to the Kodama observers.
We illustrate the derivation of Friedmann
equation, and also extend the study to the cases of Ho\v{r}ava-Lifshitz
gravity and IR modified Ho\v{r}ava-Lifshitz gravity. 
}

\received{\today} 
\accepted{\today}
\begin{document}

\section{Introduction}

Since the discovery of black hole thermodynamics\cite{Bekenstein1973,Hawking1975,Bardeen:1973gs}
in the 1970s, it has been widely accepted that a deep relationship exists
between gravity theories and thermodynamics. This relationship possibly
offers a window to the nature of quantum gravity. In 1995, Jacobson\cite{Jacobson1995}
derived Einstein equations by demanding fundamental Clausius relations
$\delta Q=T\, dS$ for all the local Rindler horizon through each
spacetime point, with $\delta Q$ and $T$ interpreted as the energy
flux and Unruh temperature seen by the accelerated observer. In the
FRW model, it is shown that the Friedmann equation which describes
the dynamics of the universe can be derived from the first law of
thermodynamics associated with the apparent horizon, by assuming
a temperature $T=\frac{1}{2\pi\tilde{r}_{A}}$ and the entropy $S=\frac{A}{4}$,
where $\tilde{r}_{A}$ and $A$ are the radius and area of the apparent
horizon\cite{Cai2005}. The derivations are proven in Einstein gravity,
Gauss-Bonnet gravity, Lovelock gravity. Later on, it is shown that
the Friedmann equation can also be regarded as the first law of thermodynamics
in scalar-tensor gravity and f(R) gravity, by properly defining the
energy density $\tilde{\rho}$ and pressure $\tilde{p}$\cite{Akbar:2006er}.
It is further pointed out in \cite{Cai2009} that the temperature
$T=\frac{1}{2\pi\tilde{r}_{A}}$ associated with the apparent horizon
can be obtained using the tunneling formalism. Thus an observer inside
the apparent horizon will see a thermal spectrum when particles tunneling
from outside the apparent horizon to inside the apparent horizon.
These tunneling particles are detected by a Kodama observer, i.e.,
the energy of the tunneling particle is defined by using of the Kodama
vector\cite{Kodama1980}. In this paper we propose a more direct way
to calculate the energy flux $\delta Q$ across the apparent horizon,
using the Kodama vector. This derivation has significant physical
meanings. Combined with the derivation of temperature in \cite{Cai2009}
, it is strongly suggested that the first law of thermodynamics associated
with the apparent horizon in the FRW model is relative to the {}``Kodama
observer''. This result enriches the discussion of thermodynamics
in FRW universe. We also suggest to calculate the energy flux across
the apparent horizon in this direct method. Then, using the temperature
$T=\frac{1}{2\pi\tilde{r}_{A}}$ corresponding with the Kodama vector
and assuming the entropy to be a quarter of apparent horizon area,
the Clausius relation $\delta Q=T\, dS$ implies the Friedmann equation
of the FRW universe model. 

Discussions on thermodynamics of the FRW universe model in various
gravities theories and brane-world scenarios also refer to \cite{cai_unified_2007,akbar_thermodynamic_2007,akbar_thermodynamic_2007-1,gong_friedmann_2007,cai_thermodynamics_2008,wu_thermodynamicsapparent_2008}.
A modified Friedmann equation can be derived from a quantum corrected
area-entropy relation\cite{Cai2008}. Recently, a new theory of gravity
at a Lifshitz point was proposed by Ho\v{r}ava\cite{Horava2009}.
It may be regarded as a UV complete candidate for general relativity.
The Ho\v{r}ava-Lifshitz theory has been intensively investigated\cite{Horava2009a,Horava2009b,Volovich2009,Cai2009a},
and its application to cosmology has been studied\cite{Calcagni2009,Takahashi2009,Kiritsis2009}.
Spherical symmetrical black hole solutions and cosmological solutions
have been found\cite{Lu2009a,Cai2009b,Cai2009c,Nastase2009}. Since
the generic IR vacuum of this theory is anti-de Sitter, in order to
get a Minkowski vacuum in the IR, one should add a term $\mu^{4}R$
into the action and take the $\Lambda_{W}\rightarrow0$ limit. Black
hole and cosmological solutions of this so called {}``IR modified
Ho\v{r}ava-Lifshitz gravity'' have also been studied\cite{Kehagias2009,Castillo2009,Myung2009a,Park2009,Peng2009,Wang2009}.
It is shown in a recent paper\cite{Cai2009d} that at the black hole
horizon in Ho\v{r}ava-Lifshitz gravity, gravitational field equation
can be casted into the form of the first law of thermodynamics. It is
quite necessary to extend the study of relations between Friedmann
equation and thermodynamics of FRW model to the cases of Ho\v{r}ava-Lifshitz
gravity. In this article, we study the cosmological solutions of Ho\v{r}ava-Lifshitz
gravity and IR modified Ho\v{r}ava-Lifshitz gravity, and show that
Friedmann equation of FRW universe model in the Ho\v{r}ava theory
can also be written in a form of first law of thermodynamics.

This paper is organized as follows. In Section \ref{sec:2} we review
the derivation of Friedmann equation from the first law of thermodynamics
in the context of Einstein gravity. We propose the direct method to
calculate the energy flux across the apparent horizon, and discuss
its physical meanings. In Section \ref{sec:3} we show that Friedmann
equation of the FRW cosmological solutions of Ho\v{r}ava and IR modified
Ho\v{r}ava gravity can also be written into the form of first law
of apparent horizon thermodynamics. Section \ref{sec:4} is for conclusions
and discussions.

\section{First law of thermodynamics and Friedmann equation\label{sec:2}}

In this section, we briefly review the first law of thermodynamics
of the FRW metric, and give our calculation of the energy flux across the apparent horizon using the Kodama vector. The homogeneous and isotropic universe model is described by the $(n+1)$
dimensional Friedmann-Lemaitre-Robertson-Walker metric\begin{equation}
ds^{2}=-dt^{2}+a^{2}(t)\,\left(\frac{dr^{2}}{1-kr^{2}}+r^{2}d\Omega_{n-1}^{2}\right)\,,\label{eq:FRW_metric}\end{equation}
 where $d\Omega_{n-1}^{2}$ denotes the line element of an (n-1)-dimensional
unit sphere and the spatial curvature constant $k=+1,\,0$ and $-1$
corresponding to a closed, flat and open universe, respectively. Defining
$\tilde{r}=a(t)\, r$, the metric (\ref{eq:FRW_metric}) can be rewritten
as\begin{equation}
ds^{2}=h_{ab}dx^{a}dx^{b}+\tilde{r}^{2}d\Omega_{n-1}^{2}\,,\label{eq:FRW_metric2}\end{equation}
 where $x^{0}=t$, $x^{1}=r$ , $h_{ab}=diag(-1,\, a^{2}/(1-kr^{2}))$.
The dynamical apparent horizon $\mathcal{H}$ is a marginally trapped
surface with vanishing expansion, determined by $h^{ab}\partial_{a}\tilde{r}\partial_{b}\tilde{r}=0$,
which gives \begin{equation}
\tilde{r}_{A}=\frac{1}{\sqrt{H^{2}+k/a^{2}}}\,,\end{equation}
 where $H\equiv\dot{a}/a$ denotes the Hubble parameter. Also notice
that the identity $\sqrt{1-kr^{2}}\biggl|_{\tilde{r}=\tilde{r}_{A}}=H\tilde{r}_{A}$
is satisfied.

Suppose that the energy-momentum tensor of the matter in the universe
has the form of a perfect fluid $T_{ab}=(\rho+p)U_{a}U_{b}+pg_{ab}$,
where $U^{a}$ denotes the four-velocity of the fluid, and $\rho$,
$p$ are the energy density and pressure, respectively. In the FRW
metric, the components of $T_{ab}$ are $T_{00}=\rho,\, T_{ij}=pg_{ij}$.
Conservation law $\nabla_{a}T^{ab}=0$ implies\begin{equation}
\dot{\rho}+nH\,(\rho+p)=0\,,\label{eq:continity}\end{equation}
 and the 00 component of the Einstein equation is the standard Friedmann
equation\begin{equation}
H^{2}+\frac{k}{a^{2}}=\frac{16\pi G}{n\,(n-1)}\rho\,,\label{eq:Friedmann_eq}\end{equation}
 which describes the dynamical evolution of the universe model.

It was proven in \cite{Cai2005} that the Friedmann equation (\ref{eq:Friedmann_eq})
can be derived from the first law of thermodynamics. They assumed that
the apparent horizon has an associated entropy and temperature, \begin{equation}
S=\frac{A}{4G}\,,\quad T=\frac{1}{2\pi\tilde{r}_{A}}\,,\label{eq:FRWST}\end{equation}
 where $A=n\Omega_{n}\tilde{r}_{A}^{n-1}$ is the area of the apparent
horizon. According to \cite{Hayward1999,Bak2000,Cai2005}, the first
law of thermodynamics is rewritten as\begin{equation}
d\mathcal{M}=A\Psi+WdV\label{eq:1stLaw}\end{equation}
where $V=\Omega_{n}\tilde{r}^{n}$ is the volume surrounded by the
apparent horizon, and\begin{equation}
\mathcal{M}=\frac{n(n-1)\Omega_{n}}{16\pi G}\tilde{r}^{n-2}(1-h^{ab}\partial_{a}\tilde{r}\partial_{b}\tilde{r})\label{eq:M-S mass}\end{equation}
is the Misner-Sharp energy\cite{Misner1964}, which can be identified
as the total energy inside the apparent horizon. Following the discussion
in\cite{Cai2005}, the work density \[
W=-\frac{1}{2}T^{ab}h_{ab}\,,\]
is regarded as the work done by a change of the apparent horizon.
The energy-supply term\[
\Psi_{a}=T_{a}^{\, b}\partial_{b}\tilde{r}+W\partial_{a}\tilde{r}\]
determines the total energy flow $\delta Q=A\Psi$ through the apparent
horizon. 

The first law of thermodynamics in FRW model has been extensively
discussed, however, a direct calculation of the energy flux $\delta Q$
from the observer viewpoint is still lacking. Here we shall propose
a simpler way of calculating the energy flux $\delta Q$ by using
the Kodama vector. It was shown in \cite{Cai2009} that there is indeed
a Hawking radiation with such a temperature (\ref{eq:FRWST}) of the
apparent horizon by using the tunneling approach. It is the first
time that the temperature of apparent horizon in FRW model can be
illustrated by another independent method. In their proof, the energy
of the particles tunneling through the apparent horizon is defined
using the Kodama vector\cite{Kodama1980}. In other words, the temperature
$T$ is {}``detected'' by a Kodama observer. This result strongly
suggests that the energy flux should be {}``detected'' by the Kodama
observer. The Kodama vector corresponding to metric (\ref{eq:FRW_metric2})
is defined as\begin{eqnarray}
K^{a} & = & -\epsilon^{ab}\nabla_{b}\tilde{r}=-\sqrt{1-kr^{2}}\left[-\left(\frac{\partial}{\partial t}\right)^{a}+Hr\left(\frac{\partial}{\partial r}\right)^{a}\right]\,,\label{eq:KodamaV}\end{eqnarray}
 where $\epsilon_{ab}=a(t)/\sqrt{1-kr^{2}}(dt)_{a}\wedge(dr)_{b}$
\cite{Jiang2009}. The Kodama vector is very similar to the Killing
vector $(\partial/\partial t)^{a}$ in the de Sitter space. In the
stationary black hole spacetime, the timelike Killing vector can be
used to define a conserved mass (energy). Since there is no timelike
Killing vector in the dynamical black hole and FRW spacetime, the
Kodama vector generates a preferred flow of time and is a dynamic analogue
of a stationary Killing vector\cite{Hayward1998}. By the Kodama vector,
a conserved quantity, Misner-Sharp energy\cite{Misner1964}, can be
defined for the FRW spacetime\cite{Hayward1996}. The use of Kodama
vector field as a preferred time evolution vector field in spherically
symmetric dynamical systems also rises simplifications\cite{Racz2006}.

Now let's calculate the energy flux through the apparent horizon.
We assume that all the energy flow across the apparent horizon is
described by the perfect fluid $T_{ab}$. Since the temperature is
{}``detected'' by the Kodama observer, the energy flux should also
be determined {}``in view of'' the Kodama observer. Referring to
the definition of energy-momentum tensor in standard general relativity
textbooks, the 4-momentum flow measured by the Kodama observer takes
the form\[
J_{a}=T_{ab}K^{b}\,.\]
Now the energy flux across the apparent horizon during an infinitesimal
time interval $dt$ is\begin{equation}
\delta Q=\int_{\mathcal{H}}J_{a}d\Sigma^{b}=\int_{\mathcal{H}}T_{ab}K^{a}d\Sigma^{b}\,.\label{eq:fluxdef}\end{equation}
Noticing that the generator (normal vector) of horizon $n^{a}=\left(\partial/\partial t\right)^{a}-Hr\left(\partial/\partial r\right)^{a}$,
the energy flux can be calculated as\begin{align}
\delta Q & =AK^{a}T_{ab}n^{b}dt\biggl|_{\tilde{r}=\tilde{r_{A}}}\nonumber \\
 & =A\left(K^{t}T_{tt}n^{t}+K^{r}T_{rr}n^{r}\right)\biggl|_{\tilde{r}=\tilde{r_{A}}}\nonumber \\
 & =A\left(\sqrt{1-kr^{2}}\rho+\sqrt{1-kr^{2}}Hr\, p\frac{a^{2}}{1-kr^{2}}\,\frac{\sqrt{1-kr^{2}}}{a}\right)\biggl|_{\tilde{r}=\tilde{r}_{A}}\nonumber \\
 & =A(\rho+p)H\tilde{r}_{A}\,\,.\label{eq:fluxcalt}\end{align}

Finally, using the expression of $T$ and $S$ in (\ref{eq:FRWST}),
the Clausius relation $\delta Q=T\, dS$ implies that\begin{equation}
TdS=\delta Q=A(\rho+p)H\tilde{r}_{A}\,.\label{eq:TdS}\end{equation}
 Noting that\[
\dot{\tilde{r}}_{A}=-H\tilde{r}_{A}^{3}\left(\dot{H}-\frac{k}{a^{2}}\right)\,,\]
 Eq.(\ref{eq:TdS}) leads to\begin{equation}
\dot{H}-\frac{k}{a^{2}}=-\frac{8\pi G}{n-1}(\rho+p)\,.\label{eq:1stFRW}\end{equation}
 Substitute $(\rho+p)$ into (\ref{eq:1stFRW}) using continuity condition
(\ref{eq:continity}), and integrate, we finally get \[
H^{2}+\frac{k}{a^{2}}=\frac{16\pi G}{n(n-1)}\rho\,,\]
 which is just the Friedmann equation, the integration constant can
be regarded as a cosmological constant and incorporated into the energy
density $\rho$ as a special component.

One comment follows. In some articles, for
example \cite{Lidsey2009,Shu2010}, the energy flux is calculated directly by using the generator
$n^{a}$ of the apparent horizon as $\delta Q=4\pi\tilde{r}_{A}^{2}T_{ab}n^{a}n^{b}dt$.
However, this expression cannot give the correct result, so we suggest
to use the expression (\ref{eq:fluxcalt}) to calculate the energy
flux $\delta Q$ in treating the apparent horizon thermodynamics in
FRW model. Compared to the previous treatment, our expression not
only gives the correct result, but also has significant physical meanings.

\section{Thermodynamics of FRW universe in (IR modified) Ho\v{r}ava-Lifshitz
gravity\label{sec:3}}

In the previous section we have briefly reviewed the equivalence between Friedmann equation and the first law of thermodynamics in FRW universe. This subject has been extensively studied in various gravity theories
besides Einstein theory. In Gauss-Bonnet gravity and Lovelock gravity
where the entropy of black holes does not obey the area formula, the corresponding Friedmann equation can also be obtained from
the apparent horizon first law of thermodynamics\cite{Cai2005},  by
employing the entropy formula of static spherically symmetric black
holes. However, in scalar-tensor gravity and f(R) gravity, in order to obtain the Friedmann equation, one still has to take the ansatz $T=1/2\pi\tilde{r}_{A}$
and $S=A/4G$, and redefine the energy $\tilde{\rho}$ and pressure
$\tilde{p}$\cite{Akbar:2006er}; it is also suggested to replace
the original Clausius relation by a non-equilibrium one , $\delta Q=TdS+Td_{i}S$
in scalar-tensor theory\cite{cai_unified_2007} and $f(R)$ theory\cite{akbar_thermodynamic_2007-1};
and it is suggested in \cite{gong_friedmann_2007} to replace the
Misner-Sharp mass by a masslike function, etc. This implies that the derivations
of Friedmann equation from first law of thermodynamics are subtle
in various gravity theories. In this section we consider the derivation
of Friedmann equation from first law of thermodynamics in the case
of Ho\v{r}ava-Lifshitz gravity. Thermodynamics of cosmological model
in Ho\v{r}ava-Lifshitz gravity has also been studied in \cite{Wang2009}\cite{Wei2010}.
Motivated by \cite{Akbar:2006er} and \cite{Wang2009}, we redefine
the energy and pressure of the perfect fluid, and show that Friedmann
equations can be casted into the form of the first law (\ref{eq:TdS}).
Here we assume the holography ansatz $S=A/4G$, and the original Clausius
relation $\delta Q=TdS$. We also study the case of IR modified Ho\v{r}ava-Lifshitz
gravity.

The recently proposed Ho\v{r}ava-Lifshitz gravity may be regarded
as a UV complete candidate for general relativity. The dynamic variables
$N$, $N_{i}$, and $g_{ij}$ are given in terms of the metric taking
the ADM form\cite{Arnowitt1962}\begin{equation}
ds^{2}=-N^{2}dt^{2}+g_{ij}\left(dx^{i}+N^{i}dt\right)\left(dx^{j}+N^{j}dt\right)\,.\label{eq:ADMform}\end{equation}
 The coordinates $(t,x^{i})$ scale differently with dynamical critical
exponent $z=3$,\[
\mathbf{x}\rightarrow b\mathbf{x},\,\, t\rightarrow b^{z}t\,.\]
 The action of Horava-Lifshitz gravity can be written as\begin{eqnarray*}
S_{HL} & = & \int dtdx^{i}\, N\sqrt{g}\left(\mathcal{L}_{0}+\mathcal{L}_{1}+\mathcal{L}_{m}\right)\,,\\
\mathcal{L}_{0} & = & \frac{2}{\kappa^{2}}\left(K_{ij}K^{ij}-\lambda K^{2}\right)+\frac{\kappa^{2}\mu^{2}\left(\Lambda_{W}R-3\Lambda_{W}^{2}\right)}{8(1-3\lambda)}\,,\\
\mathcal{L}_{1} & = & \frac{\kappa^{2}\mu^{2}(1-4\lambda)}{32(1-3\lambda)}R^2\\
 &  & -\frac{\kappa^{2}}{2w^{4}}\left(C_{ij}-\frac{\mu w^{2}}{2}R_{ij}\right)\left(C^{ij}-\frac{\mu w^{2}}{2}R^{ij}\right)\,.\end{eqnarray*}
 where the extrinsic curvature and the Cotten tensor are given by\begin{eqnarray*}
K_{ij} & = & \frac{1}{2N}\left(\dot{g}_{ij}-\nabla_{i}N_{j}-\nabla_{j}N_{i}\right)\,,\\
C^{ij} & = & \epsilon^{ikl}\nabla_{k}\left(R_{\, l}^{j}-\frac{1}{4}R\delta_{\, l}^{j}\right)\,.\end{eqnarray*}
 In the IR limit, the action should be reduced to the Einstein-Hilbert
action of general relativity\begin{equation}
S_{EH}=\frac{1}{16\pi G}\int d^{4}x\, N\sqrt{g}\left(K_{ij}K^{ij}-K^{2}+R-2\Lambda\right)\,.\label{eq:EHaction}\end{equation}
 by setting $x^{0}=ct,\,\lambda=1$, the speed of light, Newton's
constant, and the cosmological constant emerge as\begin{equation}
c=\frac{\kappa^{2}\mu}{4}\sqrt{\frac{\Lambda_{W}}{1-3\lambda}}\,,\,16\pi G=\frac{\kappa^{4}\mu}{8}\sqrt{\frac{\Lambda_{W}}{1-3\lambda}},\,\Lambda=\frac{3\kappa^{4}\mu^{2}\Lambda_{W}^{2}}{32(1-3\lambda)}\,.\label{eq:horava_CGLam}\end{equation}
 Taking the ansatz of cosmological solutions of the FRW metric form
(\ref{eq:FRW_metric}), for a perfect fluid matter contribution, the
Friedmann equations are \begin{eqnarray}
\frac{6}{\kappa^{2}}(3\lambda-1)H^{2} & = & \rho-\frac{3\kappa^{2}\mu^{2}\Lambda_{W}^{2}}{8(3\lambda-1)}\nonumber \\
 &  & +\frac{3k\kappa^{2}\mu^{2}\Lambda_{W}}{4(3\lambda-1)a^{2}}-\frac{3k^{2}\kappa^{2}\mu^{2}}{8(3\lambda-1)a^{4}}\,,\label{eq:Friedmann_horava}\\
\dot{\rho}+3H(\rho+p) & = & 0\,\,.\nonumber \end{eqnarray}

The verification of temperature $T=1/2\pi\tilde{r}_{A}$ via the tunneling
approach in \cite{Cai2009} and the derivation of the energy flux (\ref{eq:fluxcalt}) through the apparent horizon are only relevant to the FRW metric ansatz,
regardless of the specific gravity theories. It is reasonable to regard them as valid
in the context of Ho\v{r}ava-Lifshitz gravity. Furthermore, we assume
that the entropy associated with the apparent horizon is also a quarter
of horizon area, $S=A/4G$. Just as in \cite{Akbar:2006er}, and motivated
by \cite{Wang2009}, we define the energy and pressure of perfect
fluid in the universe\cite{Wang2009}\[
\tilde{\rho}\equiv\rho+\rho_{\Lambda}+\rho_{k}+\rho_{dr}\,,\,\tilde{p}\equiv p+p_{\Lambda}+p_{k}+p_{dr}\,,\]
 where\begin{eqnarray}
\rho_{\Lambda} & = & -p_{\Lambda}=-\frac{3\kappa^{2}\mu^{2}\Lambda_{W}^{2}}{8(3\lambda-1)}\,,\\
\rho_{k} & = & -3p_{k}\nonumber \\
 & = & \frac{3k}{4(3\lambda-1)a^{2}}\left(\kappa^{2}\mu^{2}\Lambda_{W}+\frac{8}{\kappa^{2}}(3\lambda-1)^{2}\right)\,.\\
\rho_{dr} & = & 3p_{dr}=-\frac{3\kappa^{2}\mu^{2}}{8(1-3\lambda)}\frac{k^{2}}{a^{4}}\,,\end{eqnarray}
 are the cosmological constant term, curvature term and the dark radiation
term. These extra terms can be viewed as the dark components, or effective
energy-momentum tensor. Identifying \begin{equation}
8\pi G_{cosmo}=\frac{\kappa^{2}}{2(3\lambda-1)}\,,\end{equation}
 and make use of the Clausius relation $\delta Q=T\, dS$, the Friedmann
equation (\ref{eq:Friedmann_horava}) in Ho\v{r}ava-Lifshitz gravity
can be obtain, \begin{eqnarray}
H^{2}+\frac{k}{a^{2}} & = & \frac{8\pi G_{cosmo}}{3}\tilde{\rho}\,,\label{eq:Friedmann_cosmo}\\
\dot{\rho}_{i}+3H(\rho_{i}+p_{i}) & = & 0\,.\nonumber \end{eqnarray}

Similar process can be made in the IR modified Ho\v{r}ava-Lifshitz
gravity. The action of IR modified Ho\v{r}ava-Lifshitz gravity is
obtained by adding a term $\mu^{4}R^{(3)}$ to the original action
\cite{Kehagias2009}, \begin{eqnarray}
S & = & \int dtd^{3}x\, N\sqrt{g}\Biggl\{\frac{2}{\kappa^{2}}\left(K_{ij}K^{ij}-\lambda K^{2}\right)-\frac{\kappa^{2}}{2w^{4}}C_{ij}C^{ij}\nonumber \\
 &  & +\frac{\kappa^{2}\mu}{2w^{2}}\epsilon^{ijk}R_{il}\nabla_{j}R_{\, k}^{l}-\frac{\kappa^{2}\mu^{2}}{8}R_{ij}R^{ij}\nonumber \\
 &  & +\frac{\kappa^{2}\mu^{2}}{8(1-3\lambda)}\left(\frac{1-4\lambda}{4}R^{2}+\Lambda_{W}R-3\Lambda_{W}^{2}\right)+\mu^{4}R\Biggr\}\,.\label{eq:IRhorava_action}\end{eqnarray}
 By taking the FRW metric ansatz and assuming the matter contribution
as perfect fluid, the Friedmann equation is\cite{Park2009} \begin{eqnarray}
H^{2} & = & \frac{\kappa^{2}}{6(3\lambda-1)}\Biggl(\rho-\frac{3\kappa^{2}\mu^{2}\Lambda_{W}^{2}}{8(3\lambda-1)}\nonumber \\
 &  & +\frac{3k\kappa^{2}\mu^{2}\Lambda_{W}}{4(3\lambda-1)a^{2}}-\frac{6k\mu^4}{a^2}-\frac{3k^2\kappa^{2}\mu^{2}}{8(3\lambda-1)a^{4}}\Biggr)\,.\label{eq:Friedmann_IRm}\end{eqnarray}
 Again, by introducing the cosmological constant term, the curvature
term and the dark radiation term\begin{eqnarray}
\rho_{\Lambda} & = & -p_{\Lambda}=-\frac{3\kappa^{2}\mu^{2}\Lambda_{W}^{2}}{8(3\lambda-1)}\,,\\
\rho_{k} & = & -3p_{k}\nonumber \\
 & = & \frac{3k}{4(3\lambda-1)a^{2}}\left(\kappa^{2}\mu^{2}\Lambda_{W}-8\mu^4\left(3\lambda-1\right)+\frac{8}{\kappa^{2}}(3\lambda-1)^{2}\right)\,,\\
\rho_{dr} & = & 3p_{dr}=\frac{3\kappa^{2}\mu^{2}}{8(3\lambda-1)}\frac{k^{2}}{a^{4}}\end{eqnarray}
 and identifying\begin{equation}
8\pi G_{cosmo}=\frac{\kappa^{2}}{2(3\lambda-1)}\,,\end{equation}
 the first law of thermodynamics $\delta Q=T\, dS$ implies the Friedmann
equation\begin{eqnarray}
H^{2}+\frac{k}{a^{2}} & = & \frac{8\pi G_{cosmo}}{3}\tilde{\rho}\,,\label{eq:Friedmann_cosmo_IR}\end{eqnarray}
 which is just Eq.(\ref{eq:Friedmann_IRm}). Thus we have shown that
Friedmann equations in Ho\v{r}ava-Lifshitz gravity and IR modified
Ho\v{r}ava-Lifshitz gravity can both be casted into the form of first
law of apparent horizon thermodynamics of FRW universe model.

\section{Discussion and conclusion\label{sec:4}}

In this article, we have verified the derivation of Friedmann equation
from the Clausius relation in the FRW universe model. The Kodama vector
plays an important role, both in the proof of the temperature associated
with the apparent horizon via the tunneling approach, and in the calculation
of the energy flux through the horizon. This indicates a physical
fact that the first law of thermodynamics associated with the apparent
horizon is relative to the {}``Kodama observer''. 
Our derivation has significant physical meanings, and enriches the discussions of thermodynamics in the FRW universe model. 
In addition, we have shown that Friedmann equations in the recently proposed Ho\v{r}ava-Lifshitz
gravity and IR modified Ho\v{r}ava-Lifshitz gravity can be casted
into the form of first law of thermodynamics. We assume the entropy
associated with the apparent horizon is a quarter of horizon area,
other than the form of spherical symmetric black hole entropy in the
cases of Gauss-Bonnet and Lovelock gravity. The new degrees of freedom
in Ho\v{r}ava-Lifshitz gravity is included in the redefined energy
and pressure. Our results are useful for further understanding of the
holographic properties of gravity theories.
\begin{acknowledgments}
We thank C.Cao, Y.J.Du, J.L.Li, and Q.Ma for useful discussions. The
work is supported in part by the NNSF of China Grant No. 90503009,
No. 10775116, and 973 Program Grant No. 2005CB724508.
\end{acknowledgments}
\bibliographystyle{JHEP}
\bibliography{FRW}

\end{document}